\documentclass[twocolumn,showpacs,amsmath,amssymb]{revtex4}
\usepackage{graphicx}
\usepackage{bm}
\begin{document}
\title{Gapless Fermi Surfaces in Superconducting CeCoIn$_5$?}
\author{Victor Barzykin} 
\affiliation{Department of Physics and Astronomy, University of Tennessee,
Knoxville, TN  37996-1200}
\author{L. P. Gor'kov}
\altaffiliation[Also at ]{L.D. Landau Institute for Theoretical Physics,
Chernogolovka, 142432, Russia}
\affiliation{National High Magnetic Field Laboratory,
Florida State University,
1800 E. Paul Dirac Dr., Tallahassee, Florida 32310 }

\begin{abstract}

According to [M.A. Tanatar \textit{et al.}, Phys. Rev. Lett. \textbf{95}, 067002 (2005)], 
in a multi-band d-wave superconductor CeCoIn$_5$ electrons remain partially uncondensed. 
Interactions must induce superconducting order on all Fermi surfaces. 
We calculate specific heat and thermal conductivity in a two band model in presence of defects. 
Superconductivity originates on one Fermi surface, inducing a smaller gap on the other. 
Impurities diminish the induced gap and increase the density of states, 
restoring rapidly the Wiedemann-Franz law for this Fermi surface. 
Our calculations are in agreement with experiment.
\end{abstract}
\vspace{0.15cm}


\maketitle

The heavy fermion "115" materials CeMIn$_5$ (M=Ir,Co,Rh) are currently under extensive
experimental scrutiny. The compounds reveal non-Fermi liquid features\cite{cedomir},
the ``two-fluid'', or, alternatively, two energy scales behavior in their thermodynamics
and magneto-transport\cite{NPF}. These materials are d-wave superconductors (SC), and
their properties in high magnetic fields testify in favor of an important role played by
the Pauli spin limitation and, more generally, the involvement of magnetic mechanism\cite{bianchi}.

More recent experiments on thermal conductivity and  specific heat in the
heavy fermion superconductor CeCoIn$_5$\cite{tanatar} have found that the 
variation of thermal conductivity with increased concentration of impurities
(Ce$_{1-x}$La$_x$CoIn$_5$) contradicts the expected behavior of nodal
quasiparticles in a d-wave superconductor\cite{graf,lee}. 
According to Ref.\cite{tanatar}, these results favor
multi-band superconductivity, with a group of light quasiparticles
that remains uncondensed and coexists with superconductivity on other Fermi surfaces (FS), even
in the absence of defects ($x=0$). It was suggested\cite{tanatar} that
such a scenario could  
explain many other experimental anomalies observed in this material, such as the saturation of 
 the penetration depth at low temperatures\cite{Ozcan}, the unexpectedly large
difference in thermal conductivity\cite{Capan} between longitudinal and transverse field orientations
in the plane with respect to the heat flow, and the apparent contradiction in the angular field dependence
of thermal conductivity \cite{Izawa} and specific heat\cite{Aoki}. 

It seems difficult to justify the idea of uncondensed electrons on separate FS in the SC state\cite{BG0}.
There are no fundamental reasons why the interactions between electrons on
any two given Fermi surfaces in a multi-band metal would become identically equal to zero. 
In other words, if the superconducting (SC) pairing takes place on one of the FS,
even an arbitrary weak interaction would inevitably induce a SC order parameter on the second FS\cite{Agt}.
At the same time, it is well-known that ordinary impurities  may lead to gapless superconductivity 
in multi-band superconductors. One example is MgB$_2$, where at least two different
Fermi surfaces participate in s-wave superconductivity\cite{mazin}. Another example of a multi-band
case is given by surface superconductivity\cite{GR,BG}, where
a 2D Fermi surface becomes split by the Rashba spin-orbit interaction\cite{rashba1,rashba2}.

The ``115'' CeMIn$_5$ materials (where $M = Rh, Ir, Co$) are multi-band
compounds. In particular, in CeCoIn$_5$ quasi-2D Fermi surfaces coexist with small 3D ones\cite{maehira,settai}. 
CeCoIn$_5$ is a d-wave superconductor\cite{Aoki}.

The main assertion made in Ref. \cite{tanatar} is that while in the d-wave  Ce$_{1-x}$La$_x$CoIn$_5$ the thermal
conductivity, $\kappa$, must obey the so-called ``universal'' law\cite{graf,lee}, experimentally it does not, and gapless carriers are
needed to explain the observed behavior.

According to the ``universal'' law, the zero-$T$ limit value of $(\kappa/T)_{T=0}$ should not
depend on the concentration of defects, $x$. The argumentation given in Ref.\cite{tanatar} has followed
the ideas of the kinetic theory. It is well known that in the absence of defects the 
density of states (DOS) at the d-node is zero at zero energy and increases linearly with energy\cite{VG,gorkov}.
Defects would introduce a finite DOS. If the latter were proportional to $x$, specific heat would be
proportional to $x$ as well. With the number of scattering centers given by the very same $x$, the concentration
of defects drops out of the new relation that connects the thermal conductivity and resistivity instead
of the familiar Wiedemann-Franz law. 

Actually, the ``universal'' law has been rigorously derived\cite{graf,lee} only in the so-called
``clean limit''. By this one means that the concentration of defects, $x$, is very small, so that one can 
neglect the changes in $T_c$ and in the value of the gap order parameter with $x$.

Another provision for the validity of the arguments given in Ref.\cite{tanatar} is that, at the same time, 
$x$ should be not too small, otherwise one
would encounter the region of a well-known singularity in DOS that appears due to the presence
of impurity scattering near $x=0$\cite{GK}, where the specific heat (or DOS) cannot be linear in $x$.
The range of applicability of this second provision is not well defined, and, as we show below,
depends on the strength of impurity centers.

The arguments given in Ref.\cite{tanatar} may look plausible in the framework of the kinetic theory, and
may even be fulfilled for the Zn-doped cuprates\cite{graf,lee,taillefer}, 
are certainly not applicable for the samples of Ce$_{1-x}$La$_x$CoIn$_5$ studied in Ref.\cite{tanatar}.
In fact, the latter are not in the ``clean limit'', as follows from Figs 1,2 of Ref.\cite{tanatar},
where at $x=0.02$ the value of $T_c$ is already significantly reduced 
\cite{tanatar}, and, hence, $\tau^{-1} \sim T_c \sim \Delta$. 
For all the above reasons
the data presented in Fig.3 of Ref. \cite{tanatar} for the thermal conductivity $\kappa(T,x)/T$ in the alloy 
Ce$_{1-x}$La$_x$CoIn$_5$ cannot be analyzed in terms of the ``universal'' law. 
Note, however, that at the lowest $x$ the data\cite{tanatar}
reveals the presence of a thermal conductivity component that rapidly decreases with the increased $x$, 
the feature reminiscent of the x-dependence for normal phase thermal conductivity, where the Wiedemann-Franz law applies. 

To accommodate this last fact, we postulate the existence of the 
second FS (FS2) with a small induced SC gap, which is therefore very sensitive to scattering on defects.
This FS, we suggest, becomes practically gapless already in the presence 
of a small amount of defects, without changing significantly SC parameters on the other, leading FS (FS1).
FS1 makes its own contributions to the specific heat and thermal conductivity. We argue that the sum of contributions
from the two FS explains the characteristic $x$-dependence for data\cite{tanatar} without invoking an unphysical idea of isolated
and uncondensed electrons on one of the FS.

In what follows we perform calculations of the specific heat and thermal conductivity for two FS separately:
first, for FS2, where one may neglect the variation of the order parameter with the concentration of defects, and, second,
for FS1, where such variations must be taken into account. The model, as it is shown below, accounts well for the main feature
in the experimental data\cite{tanatar}, at least qualitatively.

We use the standard weak-coupling BCS-like approach to unconventional superconductivity\cite{gorkov,mineev},
generalized to the multi-band situation\cite{ABG,ABGimp,zhitomirsky,golubov}. The Fermi liquid quasiparticles in bands 
$i$ and $j$ interact via a weak short-range interaction, $U_{ij}(\bm{r},\bm{r}')$. The Hamiltonian can be written as:
\begin{equation}
H_{int} = 
\frac{1}{2}\, \sum U_{ij}(\bm{q}) 
a^{\dagger}_{i \sigma \bm{k} + \bm{q}} 
a^{\dagger}_{j \sigma' \bm{k'} - \bm{q}} a_{i \sigma' \bm{k'}}   a_{j \sigma \bm{k}}.
\end{equation} 
Here $\sigma$, $\sigma'$ are spin indices, and the sum over all indices is written in momentum space, $\bm{k}$, $\bm{k'}$, $\bm{q}$.
The scattering potential for impurities has the form:
\begin{equation}
H_{imp} = \sum_{\sigma, l, j} \sum_{\bm{k}, \bm{q}}\sum_{\alpha}  
e^{i (\bm{q} - \bm{k}) \bm{R}_{\alpha}} V_{lj}(\bm{k},\bm{q}) 
a^{\dagger}_{j \sigma \bm{k}} a_{l \sigma \bm{q}},
\end{equation} 
where $l$ and $j$ are band indices and $\bm{R}_{\alpha}$ are the defect positions.
In turn, the inverse Gor'kov matrix $\hat{G}^{-1}$ defining the normal ($G$) and anomalous ($F,F^{\dagger}$) Green's function is:
\begin{equation}
\hat{G}^{-1} = \left[\begin{array}{cc}
i \tilde{\omega}_{j,n} - \epsilon_{j \bm{k}} & - \Delta_{\bm{k}} \\
- \Delta^*_{\bm{k}} &  i \tilde{\omega}_{j,n} + \epsilon_{j \bm{k}} 
\end{array} \right],
\label{one}
\end{equation}
where, as usual\cite{AG}, $i \tilde{\omega}_{j,n} \equiv i \omega_n - \Sigma_j(\omega_n)$.
In the Born approximation:
\begin{equation}
\Sigma_j(\omega_n) = n_{imp} \sum_{\bm{k}, l} \left|u_{jl}\right|^2  G_l(i \omega_n, \bm{k}),
\label{sigma1}
\end{equation}
$n_{imp}$ is the concentration of impurities, while
the self-energy for the anomalous Green's function $F_j(i \omega_n \bm{k})$ vanishes for isotropic scattering due to momentum
integration, and $|u_{jl}|^2 = \int \frac{d \Omega}{4 \pi}\, \int \frac{d \Omega'}{4 \pi}\, |V_{jl}(\bm{k}_F, \bm{k}'_F|^2$.
The Gor'kov equations are then solved, along with the gap equation,
\begin{equation}
\Delta_j(\bm{k}) = - T \sum_{\bm{q},n,l} U_{jl}(\bm{q}) F_l(i \omega_n, \bm{k}- \bm{q}).
\label{gap}
\end{equation}
We find:
\begin{equation}
\Sigma_j (\omega_n) = - \sum_l \frac{i \tilde{\omega}_{l,n}}{2 \tau_{jl}}\, \int \frac{d \Omega}{4 \pi}\, 
\frac{1}{\tilde{\omega}_{l,n}^2 + |\Delta_l(\bm{k})|^2}\,,
\label{sigma2}
\end{equation}
where
\begin{equation}
 \frac{1}{\tau_{jl}}\, = 2 \pi N_{0l}  n_{imp} |u_{jl}|^2.
\label{tau}
\end{equation}

The general analysis of the above equations when all couplings are of the same order is somewhat 
cumbersome\cite{zhitomirsky,mazin}. Neither is it necessary, since $T_c$ depends exponentially on the coupling constant,
and superconductivity will first emerge on the Fermi surface with the largest quasiparticle attraction potential (denoted
FS1 below). Even if $|U_{11}| \gg |U_{22}|, \ |U_{12}|$, 
the superconducting gap induced by the intra-band coupling on other Fermi surfaces (FS2, FS3, etc.),
although being small, has the smallness only linear (not exponential!) in $U_{12}$.
Clearly, for singlet unconventional superconductors in presence of pair-breakers 
the smaller gap will get completely ``erased'' first, resulting in a new type of
superconducting state, in which almost ``normal'' ($\tau_{22}^{-1} \gg \Delta_2$)
quasiparticles on the FS2 coexist with ``gapped'' (i.e., having the d-wave nodes) quasiparticles on the leading band.

We consider first the case when the amount of defects, while being too small to significantly change $T_c$ 
and SC parameters on FS1, 
is sufficient to change thermodynamics and transport attributable to FS2. 

The theoretical formalism is well-known (see, for instance, Ref. \cite{mineev}). However, 
numerical results are different for FS2  (compare, e.g., with Ref.\cite{SM}), 
because $\Delta$ for FS2 ($=\Delta_2$) is taken as an $x$-independent constant.

For a better readability, we provide a brief summary of the theoretical results here. In Eq.(\ref{one}) the analytical continuation 
of $i \tilde{\omega}_n$ from the upper energy half-plane defines the variable $t(E)$ (in the Born approximation) as:
\begin{eqnarray}
t(E) &=& E + \frac{i}{2 \tau}\, \int_0^{2 \pi} \frac{d \varphi}{2 \pi}\, \frac{t}{\sqrt{t^2 - |\Delta_{\bm{k}}|^2}}\,, 
\label{t}
\\
\Delta_{\bm{k}} &=& \Delta \cos{(2 \varphi)}. 
\end{eqnarray}
The density of states is:
\begin{equation}
N(E) = 2 N_0 \tau Im[t(E)],
\label{dos}
\end{equation}
and for the specific heat one has:
\begin{equation}
C(T) = \frac{1}{4T^2} \int_0^{\infty} N(E) \frac{E^2 dE}{\cosh^2\left(\frac{E}{2T}\, \right)}\,.
\label{C}
\end{equation}
Here $N_0$ is the normal DOS.
From general expressions for the thermal conductivity, $\kappa(T,x)$\cite{graf,lee}, we need only the low temperature limit\cite{lee} 
(see Ref.\cite{mineev}):
\begin{equation}
\kappa_{ii} = 
\frac{\pi^2}{3}\, N_0 v_F^2 T \int \frac{d \Omega}{4 \pi}\, \hat{k}^2_i \frac{\gamma^2}{(\gamma^2 + |\Delta_{\bm{k}}|^2)^{3/2}}\,.
\label{kappa}
\end{equation}
If in Eq.(\ref{kappa}) $\gamma \ll \Delta$, one obtains the notorious ``universal'' limit, i.e., 
$\kappa$ independent of the concentration of impurities\cite{graf,lee}:
\begin{equation}
\kappa = \frac{2 \pi}{9} N_0 v_f^2 \frac{T}{\Delta}\,
\label{unli}
\end{equation}
The formal definition of $\gamma$ is given by:
\begin{equation}
\gamma \equiv Im[ t(E \rightarrow 0) ],
\end{equation}
however, its explicit expression strongly depends on the assumption about the strength of the scattering potential for the 
defects\cite{PP}. At small $x$, in the Born approximation,
\begin{equation}
\gamma = \frac{N(0)}{2 N_0 \tau}\, = 4 \Delta e^{- \pi \tau \Delta}.
\label{born}
\end{equation}
In the opposite limit of strong scattering (the unitary limit\cite{PP}):
\begin{equation}
\gamma = \sqrt{\frac{\pi \Gamma \Delta}{\ln{\frac{32 \Delta}{\pi \Gamma}\,}}\,},
\label{unitary}
\end{equation}
where 
\begin{equation}
\Gamma = \frac{n_{imp}}{\pi N_0}\,.
\end{equation}

\begin{figure}
\includegraphics[width=3.375in]{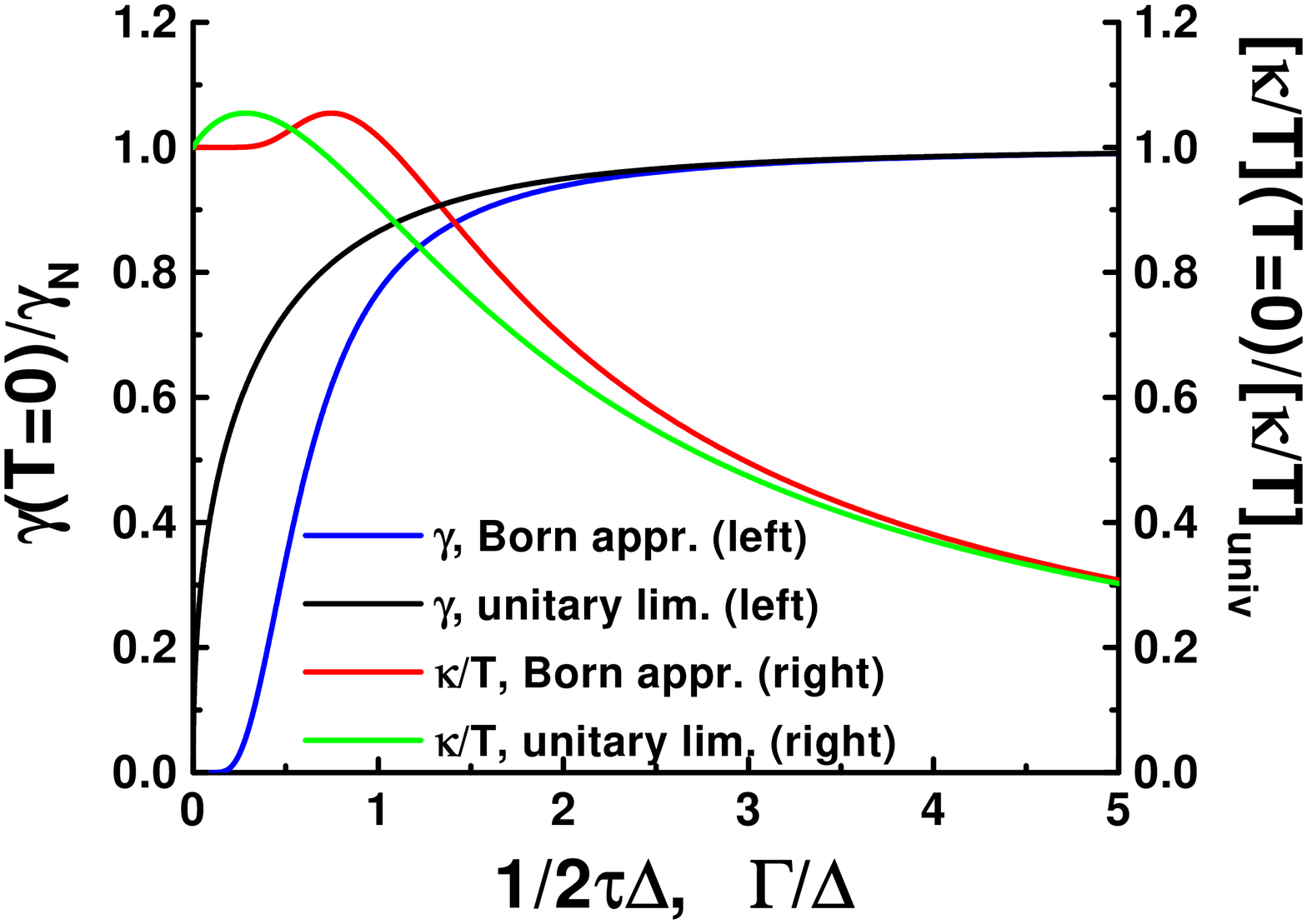}
\caption{Contribution of FS2 to the linear coefficient for the heat capacity $\gamma(T)/\gamma_N(T)$ (left axis) and
the ratio of thermal conductivity $\kappa(T)/T$ to 
its universal limit for FS2 (right axis) at small temperatures $T \rightarrow 0$ 
as a function of $1/ 2 \tau \Delta$ in
the Born approximation, or $\Gamma/\Delta$ in the unitary limit.}
\label{fig1}
\end{figure}

\begin{figure}
\includegraphics[width=3.375in]{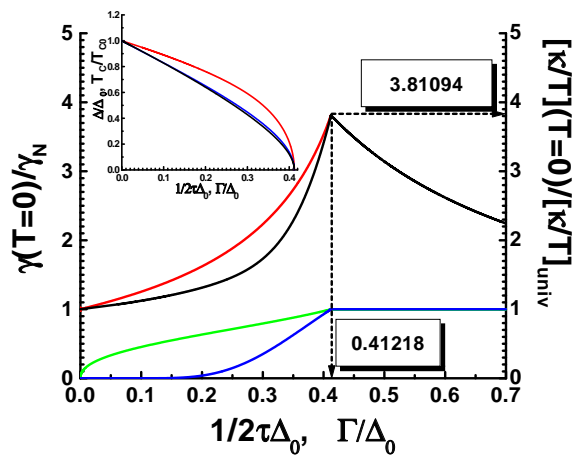}
\caption{Contribution of FS1 to the linear coefficient for the heat capacity $\gamma(T)/\gamma_N(T)$ (left axis; blue-Born approximation, green-unitary limit) and
the ratio of thermal conductivity $\kappa(T)/T$ to 
its universal limit for FS1 (right axis; black-Born approximation, red-unitary limit ) at small temperatures $T \rightarrow 0$ 
as a function of $1/ 2 \tau \Delta$ in
the Born approximation, or $\Gamma/\Delta$ in the unitary limit. The inset shows the dependence of $T_c$ (black) and the
order parameter in the Born approximation (red) and the unitary limit (blue).}
\label{fig2}
\end{figure}

The results of our calculations of the specific heat and thermal conductivity contributions from FS2 are presented in Fig.\ref{fig1}
 for these two extreme cases.
One sees that in the Born approximation a flat region (the ``universal'' law) in $\kappa/T$ appears at small $x$;
in the unitary limit for impurity scattering this interval is practically absent, although, as it was explained above,
$\Delta_2$ was kept constant. At larger $x$ $\kappa/T$ decreases, merging rapidly with the $1/x$ behavior expected from
the Wiedemann-Franz law in the normal phase. No interval is seen for the linear in $x$ behavior of the Sommerfeld
coefficient in the specific heat.

The same characteristics have been calculated for FS1 and are given in Fig.\ref{fig2}. Unlike for FS2, now both $\Delta$ and $T_c$ 
decrease
with $x$, as shown in the inset (the dependencies practically coincide with the results for the unitary limit from Ref.\cite{SM}). 
Note a considerable variation of $\kappa/T$ both in the Born approximation and in the unitary limit for
the scattering amplitude between $x=0$ and $x=x_{cr}$, where $T_c=0$ and SC is destroyed. A linear in $x$ behavior in
the specific heat is seen in some $x$-interval only in the unitary limit. Above $x_{cr}$ DOS reaches its normal value,
independent of $x$,  and $\kappa/T$  begins to decrease as $1/x$, according to the Wiedemann-Franz law,
leading to the corresponding ``bump'' at $x=x_{cr}$. Note that one sees such a ``bump'' in experimental data shown on Fig.3 
of Ref.\cite{tanatar}.

To summarize, the model of two FS with different values of SC order parameter provides an excellent qualitative interpretation
of the remarkable findings in Ref.\cite{tanatar}.
Indeed, we have shown  that at smaller $x$ the $x$-dependent contribution to $\kappa/T$ 
in Ref.\cite{tanatar} can be attributed  to a FS2 
(such as the $\epsilon$- sheet\cite{maehira,settai}) ``stripped'' of a SC gap by strong scattering. 
Linear in $x$ dependence of the Sommerfeld coefficient in the specific heat at $T=0$ seen in Ref.\cite{tanatar} may come from 
the FS1\cite{SM}.

The authors are thankful to Z. Fisk an L. Taillefer for attracting their attention to the problem and helpful discussions.
This work was supported (VB) by TAML at the University of Tennessee and (LPG) by NHFML through
the NSF Cooperative agreement No. DMR-008473 and the State of Florida.

\end{document}